\documentstyle[preprint,aps]{revtex}
\begin{document}
\draft
\title{Late-time Phase transition and the Galactic halo as
a Bose Liquid: (II) the Effect of Visible Matter}
\author{S.U. Ji and S.J. Sin}
\address{ Department of Physics, Hanyang University\\
Seoul, 133-791,  Korea}
\date{\today}
\maketitle
\begin{abstract}
In the previous work,
we investigated the rotation curves of
galaxies assuming that the dark matter consists of
ultra light boson appearing in $'$late time phase transition' theory.
Generalizing this work, we consider
the effect of visible
matter and classify the types of rotation curves as we vary the
fraction of the mass and extention of visible matter.
We show that
visible matter, in galaxies with flat rotation curves,
has mass fraction $ 2\% \sim 10\% $ and it is confined within the
distance fraction $ 10\% \sim 20\%$.
\end{abstract}

\narrowtext
\section{INTRODUCTION}
Recently, motivated by the large scale structure, an ultra light
Nambu-Goldstone boson  was
introduced in dark matter physics under the name of
"late time cosmological phase transition" \cite{hill}.
The basic idea of the theory is that if a phase transition
happened  {\it after}
decoupling, one can avoid the constraint imposed by the
isotropy of the microwave back ground.

Subsequently, one of the authors of this paper\cite{sin} tried to apply
this idea to the rotation curve of galactic haloes.
If the phase transition occurred so late,
 the universe was already big, therefore
 the relevant particle must be such that
(i) its Compton wave length provides the scale of interest
since there is no other length scale of this kind.
(ii) it should be a major component of dark matter \cite{frieman}
to be responsible for the structure formation.

If the dark matter  consists of this particle whose mass is, say, $m\sim
10^{-24}eV$ and if the dark matter
density in the galaxy is about $10^{-25}g/cm^3$,
then the inter particle distance is of order $10^{-13}cm$, while the
Compton wave length is of 10pc order.
Therefore the wave functions of the particles are entirely overlapping
and a galactic halo is a giant system of a bose liquid. In this context,
one must consider the dark matter distribution quantum
mechanically and this is done in ref.\cite{sin}.

However, there was a point that had to be clarified further in
that work. All the rotation curves obtained there were
slightly increasing. While this is consistent with most of the
data\cite{rubin}, \cite{rubin2}, there are a few galaxies
whose rotation curves are decreasing.
In fact, in \cite{sin}, the effect of the visible matter was not properly
considered. Although most of the
mass of a galaxy is attributed to the dark matter, the shape
of the rotation curve can depend on the mass profile of the
visible matter, especially near the core of a
galaxy.
In this paper, we consider the effect of the visible matter to
the shape of the rotation curve more
carefully  in the scheme of the ref.\cite{sin}.

We briefly recapitulate the main idea of the paper \cite{sin}.
The dark matter that is consist of ultra-light
boson is in a condensation state.
Usually bosons condense in the ground state, because in atomic
physics situation, matter is always coupled to electromagnetism
(cooling)  and this  efficiently lower the energy of a system.
In our case, however, it was shown \cite{sin} that
 there is no mechanism to reduce the total energy of the system, therefore
we look for the non-ground state condensation described
by a macroscopic wave function that has nodes.
We will classify the rotation curves for three thousand different
visible matter distributions.

\section{QUANTUM FLUID OF DARK MATTER}
The system is described by a macroscopic wave function
(order-parameter) $ \psi$.
For simplicity, we consider spherically symmetric case.
We also set the visible matter distribution to be spherically symmetric.
This is not realistic, because the visible matter distribution of
the disk galaxy is by definition not spherical. However, the
purpose of this paper is to show the possibility of the
decreasing curves and we believe that, for this purpose,
 it is enough to consider the spherically symmetric case only.
We set the visible matter potential by hand and get the wave function of
 total system by numerical  work.
 Also, we neglect inter-particle interactions apart from gravity.
  Since total distribution determines the potential of an individual
  particle, we expect that the system be described by
a solution of non-linear rather than linear wave equation.

Newtonian potential $V$ is given by
$$\nabla^2 V =4\pi G(\rho_{dark}+\rho_{visible})$$
and the schr\"odinger equation is
\begin {eqnarray}
 i \hbar \partial _t \psi = -\frac { \hbar ^2 }{2m} \nabla ^2 \psi
+V \cdot  \psi(r)
\end  {eqnarray}
By identifying
$\rho_{dark}=GM_0m|\psi|^2$ and defining $\rho_{visible}=GM_0m
\rho_v$
we get
\begin {eqnarray}
 i \hbar \partial _t \psi = -\frac { \hbar ^2 }{2m} \nabla ^2 \psi
+ GmM_0  \int ^{r}_{0} d \acute{r} \frac {1}{\acute{r}^2} \int ^{\acute{r}}
_{0} d r ^{\prime \prime}  4 \pi r^{\prime \prime 2} ( | \psi| ^2 + \rho _v
           )  \cdot  \psi(r)
\end  {eqnarray}
Here, $ M_0 $ is a mass parameter introduced for dimensional
reason.
We now set effective density of the visible matters
as the Plummer potential for the bulge and exponential decay for the disk.
In eq. (1) it is important to remember that
$\psi$ describes the state of the whole system
 rather than a state of an individual particle.
The justification of the non-relativistic treatment can be found
in the ref.\cite{sin}. The basic idea is $v\sim GmM\sim 10
^{-2}$, because  in our case
 $ M \sim 10 ^{12} M_\odot$ and $m\sim 10^{-24}eV$ as we shall see.

 In this paper we consider two parameters that determine visible matter
  distribution.
 The one is the mass ratio of visible matter to dark matter and
 the  other is the distance ratio of
   the distance in which $90\%$ of visible matter exist
   to that in which $90\%$ of dark matter exist.

  Now we come back to the Eq.(1).
  It is non--linear,
  hence we do not have the freedom to normalize the solution.
  However, as  shown in ref.\cite{sin}
  this equation has a scaling symmetry
  which allows us to the resolve the ambiguity due to the choice of $M_0$.
  That is, the physical quantity M does not depend on the choice of$M_0$.
   We are interested in a stationary
   solution $\psi(r,t)= e^{-iEt/ \hbar}\psi (r)$.
We simplify the Eq.(1). After scaling by
\begin{eqnarray*}
 r  = r_0 \hat {r}   \; ,\;  \psi = r_0 ^{-3/2} \hat {\psi}
 \;,\; \rho_v = {r_0} ^{-3} \hat {\rho_v} ,\;\;
 E  = \frac {\hbar ^2}{ 2m} \epsilon
      \;,\; r_0 = \frac { \hbar ^2}{2GM_0 m^2}
\end{eqnarray*}

 We can write (1) in terms of the radial wave function $ u( \hat {r}) $.
\begin {eqnarray}
\ddot {u} ( \hat {r}) + ( \epsilon    - \int ^{\hat {r} } _{0}
 d {\hat {r}}^\prime \frac {1}{{\hat {r}}^\prime}
\int ^{ {\hat {r}}^\prime  } _{0}  d {\hat {r}}^ {\prime  \prime }
 ( u^2 ( \hat {r} ^{\prime \prime}) + \hat {\rho_v}
(\hat r ^{\prime \prime }) \hat r ^{\prime \prime 2} )
  ) \cdot u(\hat {r}) =0
\end   {eqnarray}
 where  $ \hat { \psi} ( \hat {r} ) =  \frac {1}{\sqrt{4 \pi}}
 \frac{ { u}( \hat {r} ) }{\hat {r} } $.
 We solve this equation  numerically.

\section{Scaling analysis}
In equation (2) there is a scaling symmetry.  That is, the eq.(2) is
invariant  under the scaling
\begin{eqnarray}
 u \rightarrow \lambda u, \;
 \hat {r} \rightarrow {\lambda}^{-1}  \hat{r},\;
 \epsilon \rightarrow {\lambda}^2    \epsilon  ,\;
 {\hat \rho }_v     \rightarrow   {\lambda}^4   {\hat \rho }_v      .
\end {eqnarray}
Therefore, if $u(\hat{r},\epsilon, {\hat \rho_v} )$ is a solution,
           so is $ \lambda u(\lambda \hat{r}, \lambda^2 \epsilon
	           ,\lambda^4 {\hat \rho_v} ) $.
 Let $\hat{M} = {\hat M}_{dark} + {\hat M}_{visible}
              = \int u^2 d\hat{r} +\int \hat {\rho} {\hat {r}}^2  $.
 When we sacale    $ \hat {r}\; \rightarrow \; \lambda ^{-1} \hat{r} $,
     $ \hat {M}\; \rightarrow \; \lambda \hat{M}$.
Since $r= r_0 \hat {r}$
is a physical quantity, it should be invariant under the
scaling. Therefore $ r_0 \to \lambda ^{-1} {r_0} $.
On the other hand, $r_0 \; \sim \;  1/(M_0 m^2 )$, hence
$M_0 m^2 \to \lambda M_0 m^2 $. The scaling of the normalized
velocity $\hat v$ can be found as follows.
The virial theorem tell us $ v^2= GM/r$, so
$v=\sqrt{GM_0/r_0}\sqrt{\hat{M}/ \hat{r}}:=v_0\cdot{\hat v}$
Therefore under the scaling above,
we should get $v_0 \to \lambda^{-1} {v_0} $ to get
physical velocity $v_0\hat{v} $. On the other hand,
$ v_0 \; \sim \; M_0 m $. Hence we get $ M_0 m \to \lambda ^{-1}
M_0m$. Summarizing, under the scaling
 $ {\hat r } \;\rightarrow\; \lambda^{-1}  {\hat r} $,
  \begin {eqnarray*}
  M_0 m^2 \to \lambda ^{-1}M_0 m^2  \\
  M_0 m   \to \lambda ^{-1}M_0 m
 \end   {eqnarray*}
Therefore $m$ is scaling invariant while $M_0 \to
\lambda^{-1}M_0 $. Consequently, $ M = M_0 \cdot {\hat M} $
is  scaling invariant  as it should be.

Why did we do above analysis?
In  solving Eq.(2), we must choose the two initial values.
 The one is  chosen   ${\hat u(0)=0}$, because ${\hat u(r)}$ is a
  radial wave function.
In case of linear Sch\"odinger equation, there is a freedom of
normalization so that one can freely choose first derivative of
$u$ at $r=0$, or $\psi(0)$. Here, the equation is non-linear and
we do not have the freedom of normalization. We are in trouble
unless something happens.
It is the scaling symmetry we proved that save our trouble.
In our numerical work,
    whichever we choose ${\hat u}_0$ or $\lambda{\hat u}_0$ as the
     initial  value, the physical quantity is not changed.
 We set $ \hat {u}(h)={0.05}*h $, where $h$ is
  a small segment of ${\hat r}$.

 Now, we set the visible matter distribution as follows.
 \begin {eqnarray*}
{\hat \rho}_v{\hat r} = \left\{
\begin{array}{ll}
{\hat \rho}_0 \cdot \frac {1}{ (( {\hat r}/{\hat r}_d)^2 +1)^2.5}  &
	   	    { \rm\;\; for\;\; inner\;\; bulge}\\
	   {\hat \rho}_v ({\hat r}_d) \cdot e^{({\hat r} _d-{\hat r} )} &
	            { \rm\;\; for\;\; the \;\;disk}
 	 \end {array}
    \right.
    \end {eqnarray*}
% In this condition velocity curve is very similar to fig-2 in chapter II.
 We study for  various values of  $\rho_0$ and ${\hat r}_d $.
We classify the rotation curve ($v(r)$) with the mass and the
distance ratio of the visible matter to the dark matter.

\section{COMPARISON WITH OBSERVED DATA}
Now we present the main result of this paper.
We looked for solutions with 5 or 6 nodes.
 This choice is motivated partly because,
 in this state there is a rotation curve very similar to NGC2998, and NGC801.
  Additional reason is if we take node number bigger than 7,
  the total mass of the halo is larger than $ 10^{13} M _\odot  $.
  Hence there are some restrictions in our choice.
  Furthermore, in node number 4,3,2,1,  the flat region is  very narrow,
  and very hard to classify. So, we choose node number 5, and 6.

 In figure-1, we  compare NGC2998 and
 5.6M--19.7D where 5.6M means the ratio of (visible matter mass)
 / (total mass) is 5.6\% and  19.7D means
(90\% of visible matter dispersion distance )
 / (90\% of total matter dispersion  distance) is 19.7\%.
 There are four ripples  in the NGC2998's data. But  we suspect that outside
  the measured region there might still be density peaks and nodes that could
  lead to further ripples.
We can see that
the shapes of measured and computed  rotation curves are very similar
except near the core of the galaxy, where
the error of measured data is relatively big.
By comparing the maximal velocity and the size of a ripple, we
can get $v_0$ and $r_0$. By using $m= \hbar /( \sqrt{2} r_0 v_0) $ and $ M_0
     = r_0 {v_0}^2 / G $,
we get $m=2.90 \times 10^{-24} eV$
and  $ M_0 = 7.117 \times 10 ^{11} M_\odot $ in this case.

In figure-2, we  compare NGC801 and 6.9M--12.7d.
 In this case $m=3.14 \times 10^{-24} eV$
 	and   $M_0 = 6.54 \times 10 ^{11} M _\odot$.

 In the figures-3 and 4, Y axis is visible matter's
 $'$mass ratio' to total mass,
 and X axis is $90\%$ of the visible matter $'$dispersion distance'
  to the total distance.
 The figures 3, 4 are the
  classification of the shapes of rotation curves.
 The symbol / means that corresponding
rotation curve is increasing type, similarly
$\backslash$ means decreasing, and -- means flat.
 When we measure a rotation curve in a real galaxy,
 we can't see full shape since we can observe only
to the extent that sufficient amount of the hydrogen gas exists.
 So  up, down,  flat (/, $\backslash$ , -- )
 classification is not for the full shape,
but for the region that corresponds to first four or five ripples.
The figures show that in a galaxy with relatively big visible
mass ratio and small extension ratio has decreasing rotation
curves, while the one with small mass ratio and large extension
ratio has increasing data. This result is entirely consistent
with the observation that compact bright galaxies have
decreasing rotation curves and small dwarf galaxies have
increasing rotation curves\cite{ca}.

To give more intuition in reading the figures 3 and 4, we draw three
rotation curves with the same mass ratio and different distance
ratios. See figures 5, 6, 7, 8. We also give rotation curves with
the same distance ratio and different mass distance ratios.

Notice that
we can find some interesting fact on the  visible matter distribution
from  figures 3, 4.
Our analysis shows that the flat region is confined in a
small window, mass ratio 2\% to 10 \% and distance ratio 10\% to 20\%.
On the other hand,
the overall flatness is a general phenomenon in the real galaxies.
Consequently, in our frame work, we can say that most galaxies, which
have flat rotation curves, should have the mass and distance ratio in
this range.

In figure 5, we fix the mass ratio $ 6.5 \% \pm 0.5\% $, and
    change the rate of dispersion distance. Figure 6 is node 6 case.
 As  we change the mass fraction of the visible
  matter, we can find three different types of rotation curve.
 With the change of the mass fraction, we can obtain various types of rotation
 curves.
If the visible matter distribution is highly confined, the shape
of the graph
    is  of a decreasing form.
 When the visible matter distribution is widely dispersed,
    the graph is of slightly increasing a form that is very similar to
    the case of dark matter only.
 When its distribution is neither very  dispersed nor very  condense,

 the graph is almost flat.

In the case shown in figures 5,6, the mass fraction of visible matter
   is not so large. So, the shape of
    the graph is not so different to the case of dark matter only.
 However, when the mass fraction of visible matter is  very large,
    the  situation is quite different.  See figures 7, 8.
In figure 7, we fix the mass ratio $ 25\% \pm 2\% $, and
change the distance ratio in node 5. The same analysis for node
6 is shown in figure 8.
Notice that in figures 7, 8 there is no flat curve in any case,
 and the shapes are quite different to the case of dark matter only.

In  figure 9,  we fix the distance ratio $ 12\% \pm 1\% $, and
    change the  mass ratio node 5 case. Figure 10, is for node 6 case.

\section{DISCUSSION AND CONCLUSION }

In this paper we investigated the rotation curves of
galaxies assuming that the dark matter consists of
 ultra light boson appearing in late time phase transition.
Galactic halos made  of this species are highly correlated bose liquid.
We develop a Landau-Ginzberg type theory describing the collective
 behavior of the system.

As the main results, we can obtain decreasing curves as well as
flat and slightly increasing curves.
In the previous work \cite{sin}
we could see only slightly increasing curves, while observed
data shows that a few galaxies have decreasing rotation curves.
Here, we can see the flat curves and the decreasing curves as well
by considering the visible matter, in addition to the dark matter.
We show that the small amount of dark matter can change the shape of
rotation curves near the core significantly.

Our paper has one interesting  prediction.
The flat case is confined in a rather small window
corresponding to the mass ratio $ 2\% \sim 10 \%$
    and in the distance ratio $ 10\% \sim 20\%$.
On the other hand of the observed galaxies has a 'flat rotation curve'
So, we can predict that in most of galaxies
have the mass ratio and distance ratio in that region.

So we conclude that the ultra light boson appearing in the late
time phase transition theory has good enough properties to be a
dark matter candidate, at least from the rotation curve point of view.

\noindent{\bf acknowledgement}
This work was supported by research fund of Hanyang University.

\newpage
%\center{\bf \Large Figure Captions}
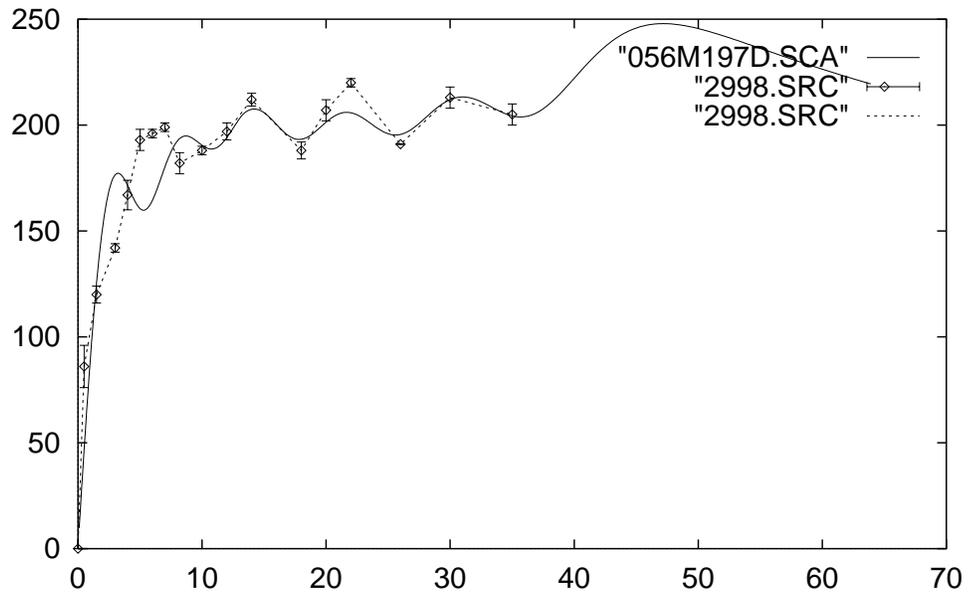
\begin{figure}
 %\vspace {65mm}
 \caption { comparing the theory and observation for NGC2998 with the data
 that $5.6\%$ of mass fraction and $19.7\%$ of distance fraction}
 \end {figure}

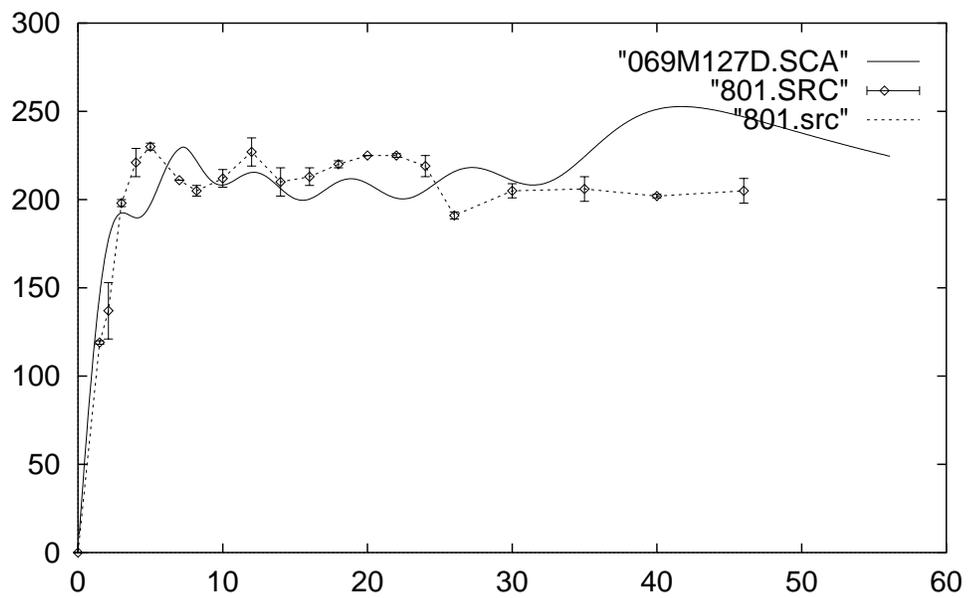
\begin {figure}
%\vspace {65mm}
\caption {comparing the theory and observation for NGC801 with the data
 that $6.9\% $of mass ratio and $12.7\%$ of distance ratio}
\end {figure}

%\newpage
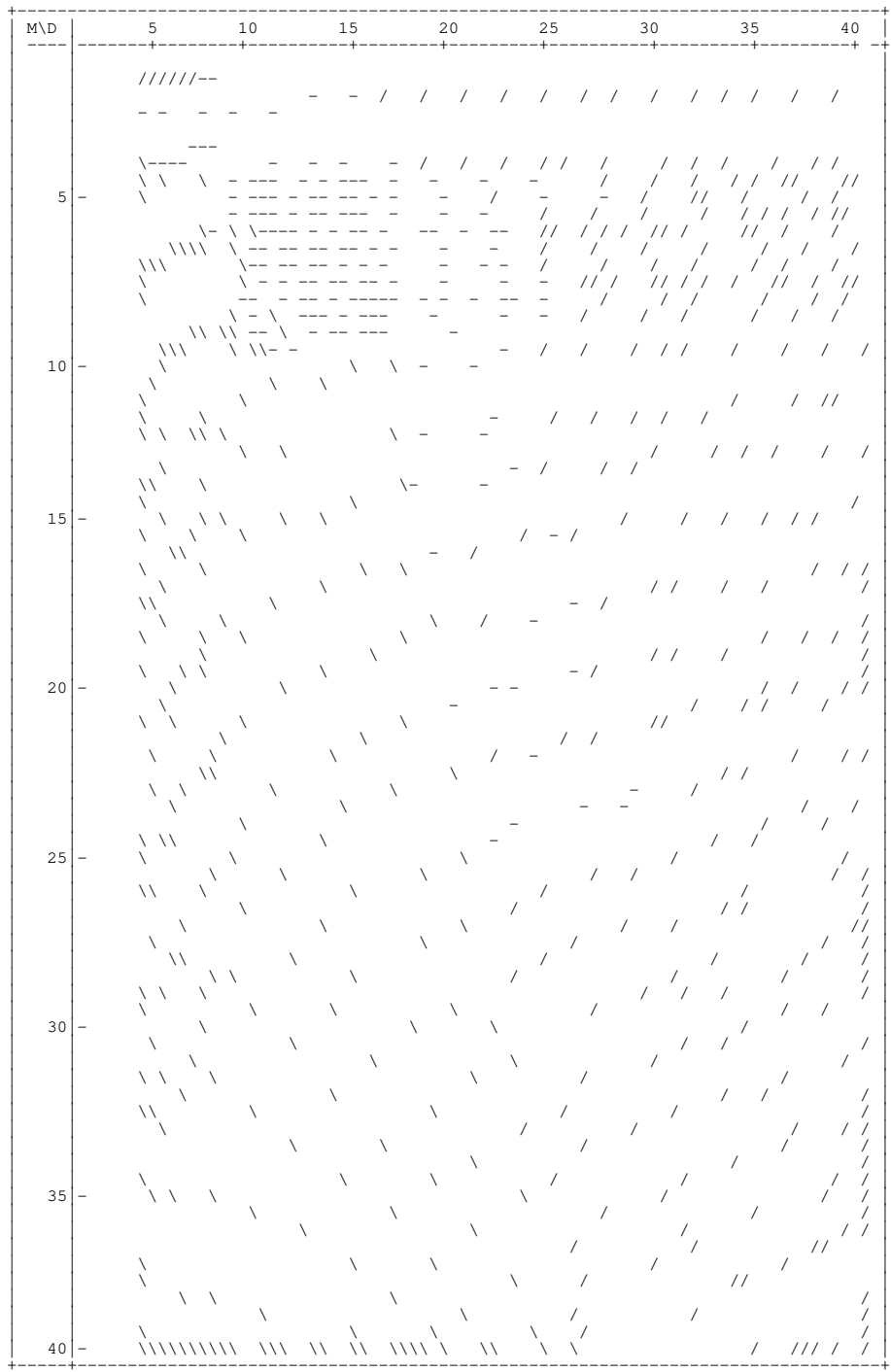
\begin   {figure}
%\vspace  {150mm}
\caption {classification of node 5 data; / means increasing,
           --- means flat, $\backslash$ means decreasing rotation curve \hfill}
\end {figure}

%\newpage
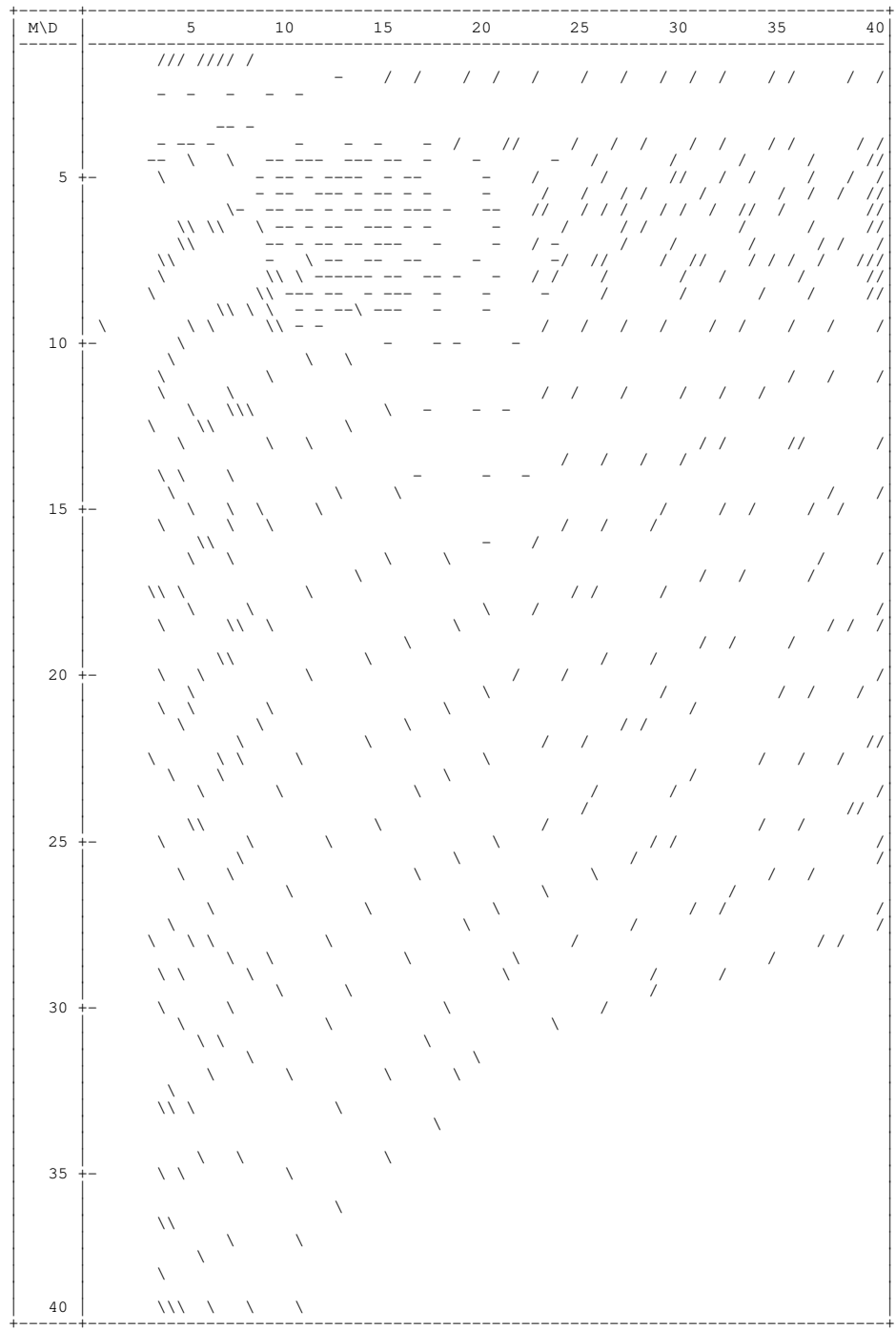
\begin   {figure}
%\vspace  {150mm}
\caption {classification of at node 6 data}
\end {figure}

%\newpage
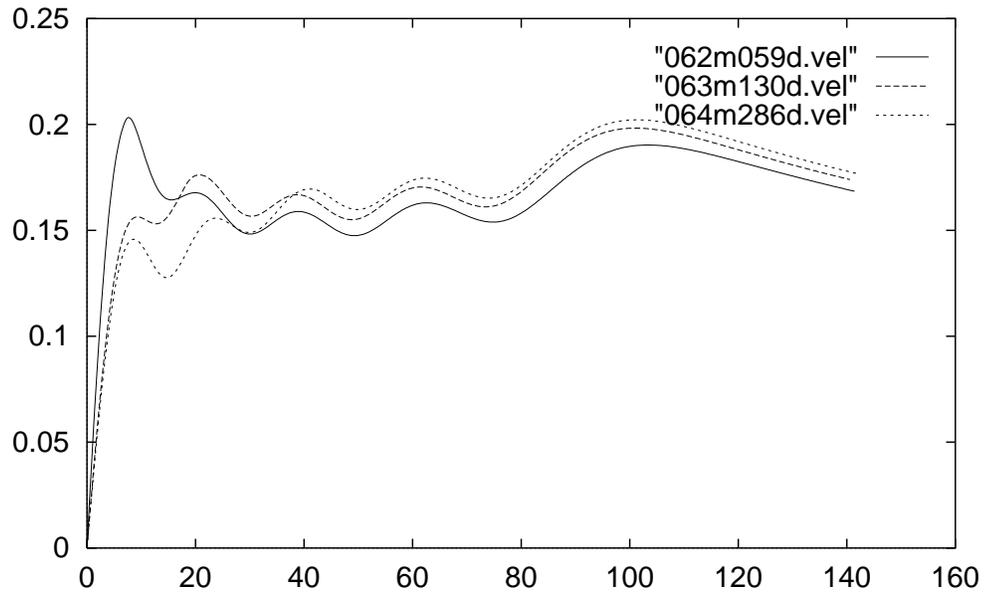
\begin {figure}
 %\vspace  {65mm}
 \caption {(1) mass ratio $ = 6.4\%$     and distance ratio  $=28.0\%$.
	   (2) mass ratio $ = 6.3\%$     and distance ratio  $=13.0\%$.
	   (3) mass ratio $ = 6.2\%$     and distance ratio  $=5.9\%$.
	   AT node 5 }
\end {figure}

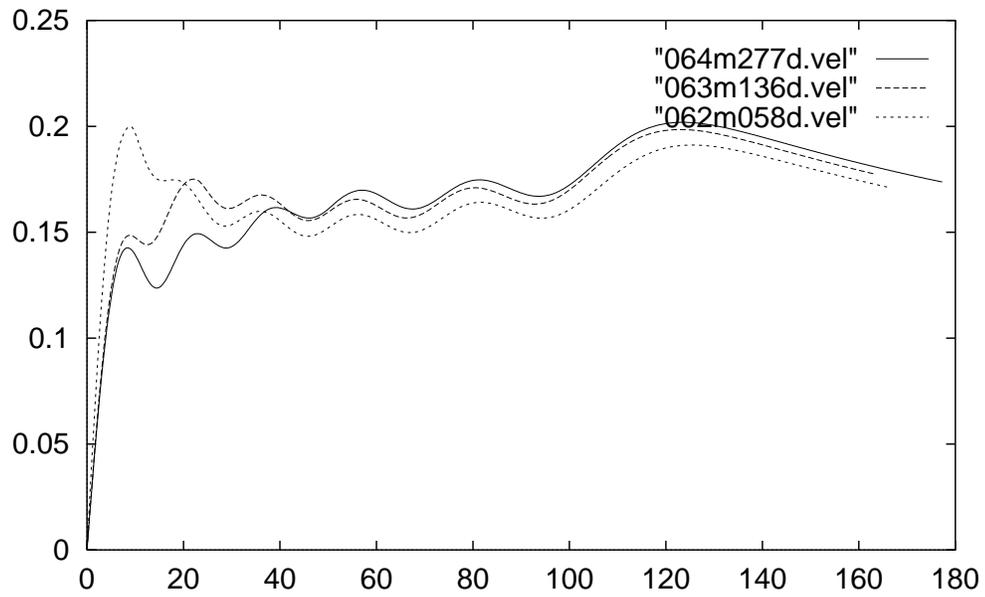
\begin {figure}[b]
 %\vspace  {65mm}
 \caption {(1) mass ratio $ = 6.4\% $ and distance ratio  $=27.7\%$.
	   (2) mass ratio $ = 6.3\% $ and distance ratio  $=13.6\%$.
	   (3) mass ratio $ = 6.2\% $ and distance ratio  $=5.8 \%$.
   		At node 6}
\end {figure}

%\newpage
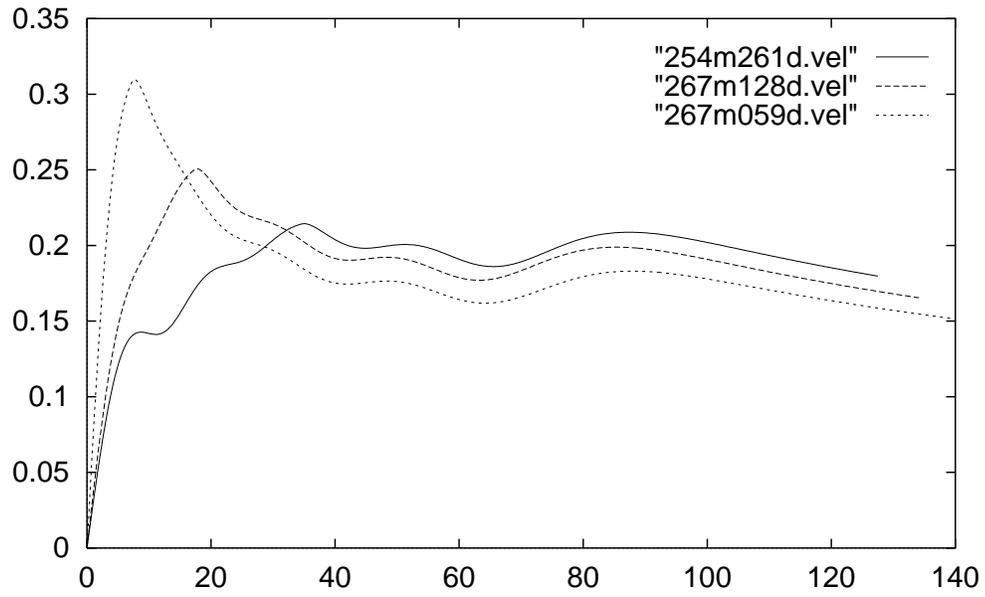
\begin {figure}
 %\vspace  {65mm}
 \caption {(1) mass ratio $ = 25.4\% $ and distance ratio  $=27.4\%$.
	   (2) mass ratio $ = 26.8\% $ and distance ratio  $=13.8\%$.
	   (3) mass ratio $ = 26.8\% $ and distance ratio  $=5.9\%$.
   At node 5}
 \end {figure}

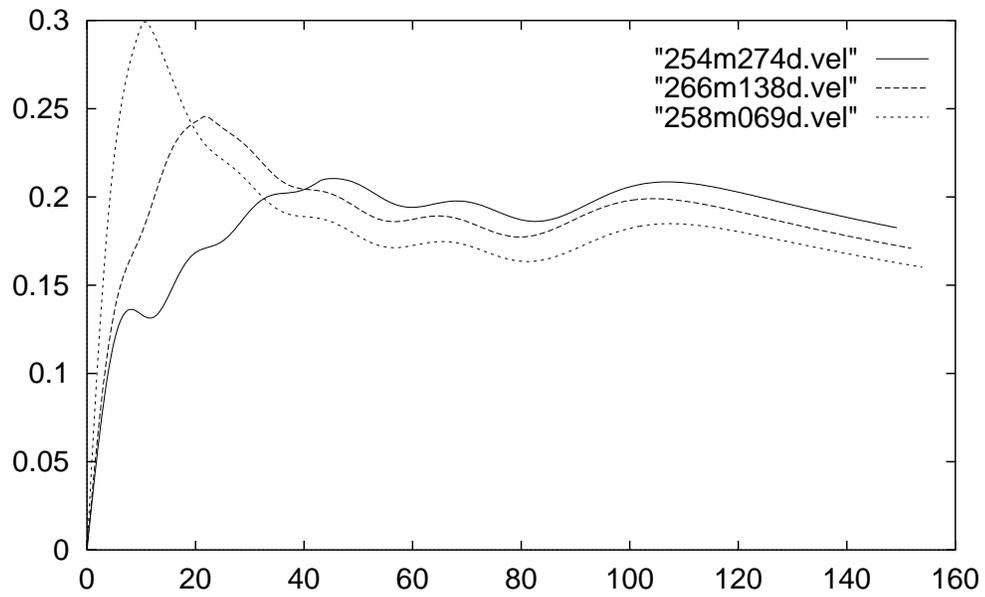
\begin {figure}[b]
 %\vspace  {65mm}
 \caption {(1) mass ratio $ = 25.4\% $ and distance ratio  $=26.1\%$.
	   (2) mass ratio $ = 25.8\% $ and distance ratio  $=14.3\%$.
	   (3) mass ratio $ = 25.8\% $ and distance ratio  $=6.9\%$.
   At node 6}
 \end {figure}

%\newpage
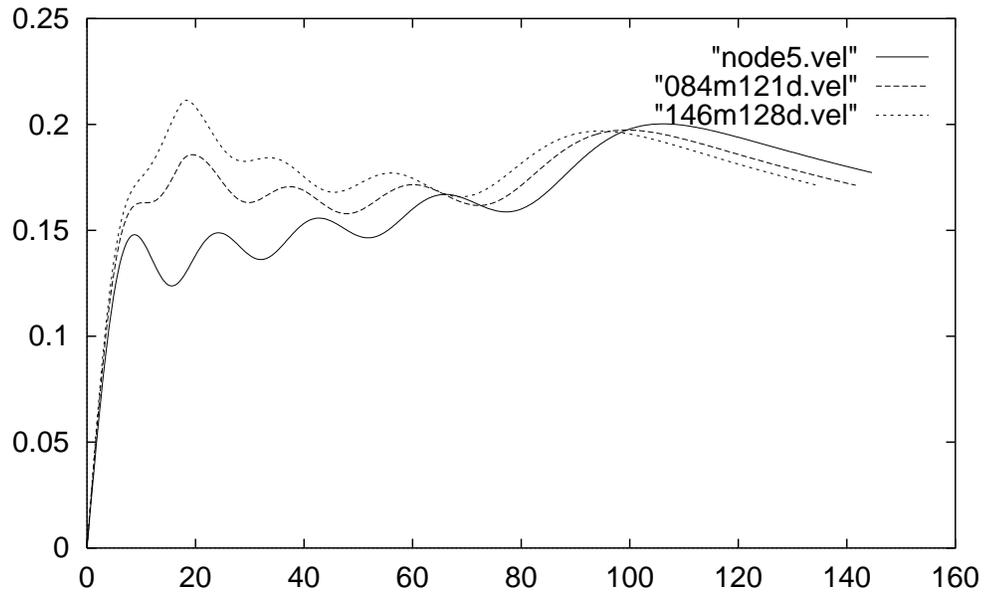
 \begin {figure}
 %\vspace  {65mm}
 \caption {(1) mass ratio $ = 0\% $.
	   (2) mass ratio $ = 8.4\% $ and distance ratio   $=12.1\%$.
	   (3) mass ratio $ = 14.6\% $ and distance ratio  $=12.8\%$.
   At node 5}
 \end {figure}

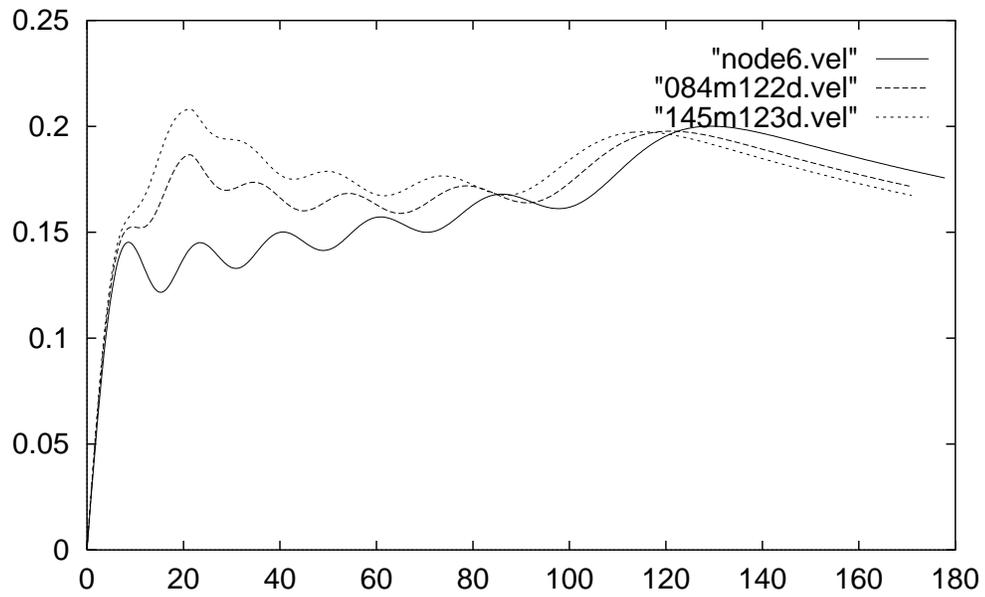
\begin {figure}
 %\vspace  {65mm}
 \caption {(1) mass ratio $ = 0 \% $.
	   (2) mass ratio $ = 8.4\% $ and distance ratio   $=12.2\%$.
	   (3) mass ratio $ = 14.5\% $ and distance ratio  $=12.3\%$.
   At node 6}
 \end {figure}

\end{document}